\documentstyle[12pt,a4,psfig]{article}

\newcommand{\be}{\begin{equation}}
\newcommand{\ee}{\end{equation}}

\def \g_ib{{\cal G}(i,b)}
\def \fij{f_{ij}}
\def \fsij{\fij^{\hbox{\scriptsize \bf self}}}
\def \flij{\fij^{\hbox{\scriptsize \bf load}}}
\def \eij{\epsilon_{ij}}
\def \kij{k_{ij}}

\begin{document}
\leftline{\small Proceedings of \it Rigidity Theory and Applications}
\leftline{\small Traverse City, MI, June 14-18 1998}
\leftline{\small Fundamental Material Science Series, Plenum}
\vspace*{8ex}
\leftline{\large \bf GRANULAR MATTER INSTABILITY: A STRUCTURAL}
\vspace*{2ex}
\leftline{\large \bf RIGIDITY POINT OF VIEW}
\vspace*{6ex}
\leftline{\hspace{1in}Cristian F.~Moukarzel}
\vspace*{2ex}
\leftline{\hspace{1in}Instituto de F\'\i sica,}
\leftline{\hspace{1in}Universidade Federal Fluminense,}
\leftline{\hspace{1in}24210-340 Niter\'oi RJ, Brazil.}
\leftline{\hspace{1in}email: \bf cristian@if.uff.br}
\vspace*{6ex}
\section{INTRODUCTION}
Granular materials are ubiquitous in nature and  very common in industrial
processes, but it is only recently that their unusual properties have begun to
receive detailed attention from the physicists community~\cite{ReviewJNB,FACD}.
The earliest documented studies of granular matter date back to
Faraday~\cite{Faraday}, who discovered  
the convective behavior of vibrated sand, and Reynolds~\cite{Reynolds}, who
noted that compactified granular matter cannot undergo shear without
increasing its volume. 

The behavior of vibrated granular matter in some aspects resembles that of a
fluid, although there are crucial differences. Size
segregation~\cite{Sizeseg}, for example, at first sight defies intuition. When
a mixture of particles of different sizes is subject to vibration, the larger
ones  migrate to the top, irrespective of density.  Also interesting is the
layering instability~\cite{Makse} of a binary mixture under pouring. Instead
of a homogeneously mixed pile, under certain conditions an
alternation of layers of both kinds of particles can be obtained.

Similar demixing phenomena occur in granular materials subject to
various kinds of external excitation. These seem to contradict the naive
expectation that shaking should favor mixing, or take the system to a
low-energy state.  Many of the unusual properties of vibrated granular
matter are in fact due to the dissipative character of interparticle
collisions. An interesting example of the consequences of dissipation is
inelastic clustering~\cite{Clustering}, by which  particles tend to
cluster together as their relative kinetic energy is completely lost during
collisions.

The compactification of vibrated sand has been recently found to be
logarithmically slow~\cite{Knight}, resembling glassy behavior, and a spin model
with frustration has been proposed to model this process~\cite{Compactification}.
This  provides a bridge between the dynamics of spin glasses and vibrated
granular matter.

It is thus clear that granular materials present extremely interesting
dynamic phenomena, but it is already at the much simpler level of static 
properties that unusual behaviors show up. Stress propagation in
piles or packings of granulate matter has many uncommon features. 
When grains are held in a tall vertical silo, for example, the
pressure at the bottom does not indefinitely increase with height but
saturates after a certain value~\cite{Silos}. The excess weight is deviated
towards the walls and equilibrated by friction forces. A related phenomenon is
the 
formation of a pressure ``dip'' right below the apex of a conical pile of
granular matter~\cite{Dip}, instead of the expected pressure maximum. 
These phenomena indicate that gravity-induced stresses do not propagate
vertically but often deviate laterally. Pressure saturation in silos and
the pressure dip under piles are both due to the formation of
``arches''~\cite{FACD,Edwards,archmodel}. Many proposals to explain
arching~\cite{Edwards,archmodel} rest on the idea that friction plays an
essential role, but recent studies~\cite{OR,Ugo,Luding,tbp} show that friction
is not necessary. 

Photoelastic visualization experiments~\cite{Dantu,Travers,Miller,Liu} show
that stresses in granular matter concentrate along well defined paths. It is
not clear whether the characteristic size of these patterns is finite, or
limited by system size only.
Stress-concentration paths are observable even on regular packings of
monodisperse particles~\cite{Dantu,Travers}, their exact location being
sensitively 
dependent on very weak particle irregularities.
Stress-paths often suffer sudden rearrangement on a global scale
when the load conditions are slightly changed~\cite{Dantu,Miller,Radjai}.
For similar reasons, the fraction of the total weight that reaches the base of
a silo 
can vary by large amounts under very weak perturbations, or when repeating the
filling procedure with exactly the same amount of
grain~\cite{archmodel}. These phenomena demonstrate that slight perturbations
can produce macroscopic internal rearrangements in granular matter. In other
words, granular matter are internally \emph{unstable}.

In part because of the technological importance of the problem, and also
because of its interest from the point of view of basic science, much work has
been done in recent years to understand the propagation of stresses in granular
systems. On the numerical side, several methods have been implemented.
Classical Molecular Dynamics simulations~\cite{Luding,Liu,RSH-M,qmodel}, which 
usually include a fictitious damping term in order to allow the system to come
to rest, are normally very cpu-intensive and thus limited to relatively small
sizes.  
Alternatively, the elastic equations can be solved using symbolic software in
order to obtain stress values which are free of numerical error~\cite{Oron}.
Lattice automata based on random contact disorder~\cite{Jan}, are able to
reproduce the observed dip under granular piles. Contact Dynamics
simulations~\cite{Radjai} provide an efficient way to include friction forces,
and allowed numerical visualizations of stress concentration on relatively
large systems. Lubricated Granular Dynamics~\cite{OR} is a method to obtain the
equilibrium contact network of infinitely stiff networks and is based on the
use of a fictitious damping with a singularity at zero distance.  

A large number of theoretical approaches to this problem are formulated on
a continuum~\cite{Edwards,Cont-1,Cont-2,Cont-3,Cont-4} and thus rest on the
assumption that a length scale exists, below which fluctuations are negligible
when compared to averages. It is not clear whether this assumption is easily
justified for granular matter, where stress fluctuations seem to be at least
as large as average stresses~\cite{Miller,Liu}.

A different type of modeling strategy starts by formulating a stochastic
rule for stress propagation on a lattice, which is thereafter solved by
various methods~\cite{Liu,qmodel,vmodel,Socolar,EC}, or taken as the starting
point for a continuum description~\cite{vmodel}. In the simplest
version of this approach~\cite{Liu,qmodel}, only the vertical component of the
transmitted force is considered, i.e. the problem is reduced to a scalar one.  
Despite the roughness of the  approximation, this procedure gives good
results for the average distribution of stresses $P(w)$, in particular the
observed exponential decay for large stresses~\cite{Miller,Liu,Radjai}.
The occurrence of small stress is though strongly underestimated within
this simple scalar model~\cite{Liu,vmodel}. This is due to the fact that
scalar ``stresses'' propagate vertically, with at most a diffusive
width due to disorder, and therefore arching is not possible~\cite{vmodel}. 
In order to correctly reproduce the small-stress part of $P(w)$, which is
arch-dominated, the vectorial nature of stresses has to be taken into
account~\cite{vmodel,Socolar,EC}. This brings in the problem of stress signs,
since now negative (traction) stresses, which do not exist on non-cohesive
granular matter, cannot be easily avoided~\cite{vmodel}. \\
\begin{figure}[htb] 
\centerline{\psfig{figure=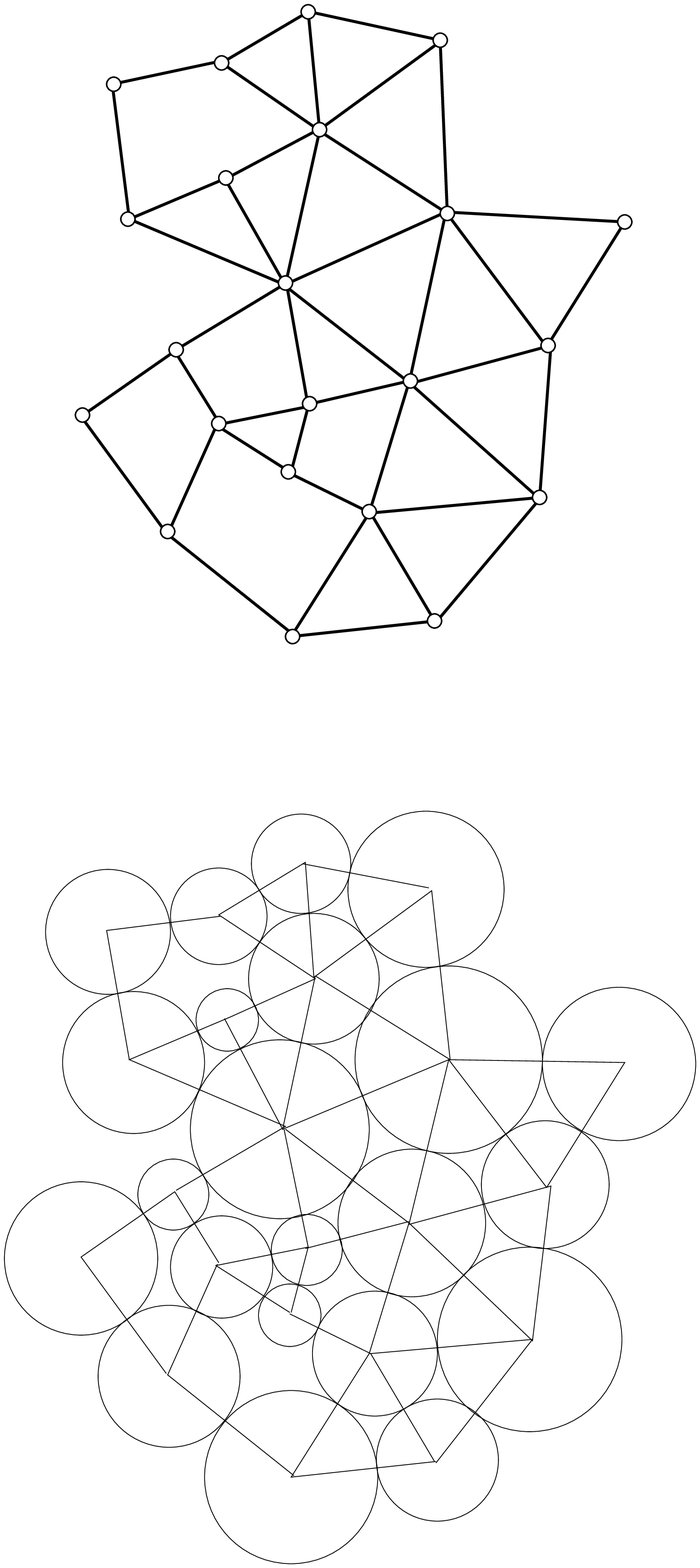,width=12cm,angle=270}}
\caption{ The contact network (right) associated to a granular pile is
a graph in which nodes represent particles and are connected by an 
edge or bond whenever there is a nonzero (compressive) force between the
corresponding particles.} 
\label{fig:contact-network} 
\end{figure} 

It is thus clear that granular matter has, from the point of view of its
static properties, two noteworthy characteristics:
\begin{itemize}
\item Stresses are not homogeneously distributed over the system
but concentrate on  paths that form  a sparse network.
\item The exact location of stress paths is susceptible of change
under very weak perturbations, showing that granular matter is extremely
unstable.  
\end{itemize}

Although there have been many proposals to describe stresses in  granular
matter,  most of these models are largely phenomenological in nature and
sometimes contain unclear ad hoc assumptions. A deeper understanding of the
above described particularities of granular matter has 
remained elusive. It is the purpose of this work to show that structural
rigidity concepts can help us advance in this direction. 

We first review some structural rigidity notions in Section~\ref{sec:SR},
and demonstrate in Section~\ref{sec:isostaticity} that the contact network of
a granular system becomes \emph{exactly isostatic} in the limit of large
stiffness. The consequences of this are discussed in
Section~\ref{sec:consequences}. An immediate consequence of isostaticity is the
possibility of stress concentration, as briefly discussed in the beginning of 
Section~\ref{sec:consequences}. Most of the previous theoretical and numerical
effort has concentrated on the description of stresses. There is though
a complementary aspect to this problem, which has not been explored. This
is the study of how \emph{displacements} induced by a perturbation propagate
in a granular system. We thus leave further discussion of stresses for
future publications~\cite{tbp}, and concentrate in understanding the
behavior of induced displacements upon perturbation. This will lead us to the  
central results of this work, respectively: \\
\noindent {\bf a)} in Section~\ref{sec:susceptibility}
it is shown that an \emph{isostatic phase transition} takes place in the limit
of infinite stiffness, and that the isostatic phase is characterized by an
anomalously large susceptibility to perturbation. \\
\noindent {\bf b)} Section~\ref{sec:response} contains a discussion of  the load-stress
response function of the system in the light of {\bf a)}, which shows that 
isostaticity is responsible for the observed instability of granular matter.

We will furthermore find that very large displacements are produced on
isostatic networks when a site is perturbed. Section~\ref{sec:pantographs}
clarifies the origin of these anomalously large displacements, while
Section~\ref{sec:conclusions} contains our conclusions.

\section{STRUCTURAL RIGIDITY AND GRANULAR NETWORKS}
\label{sec:SR}

The contact network of a frictionless packing of spherical particles can be
defined in the following way (Fig.~\ref{fig:contact-network}): we let each
particle center be represented by a 
point in space, and connect two of these points by a line (bond, or edge) whenever
there is a nonzero compression force between the corresponding particles. 
The networks so generated can be seen as particular cases of what is usually
called \emph{frameworks} in rigidity theory, i.e. structures made of points
connected by rotatable bars.  \\
Structural rigidity~\cite{Crapo,Servatius} studies the conditions that a
network of points connected by central forces has to fulfill in order to
support applied loads, i.e. be \emph{rigid}.  
The first studies of rigidity of structures from a topological point of
view date back to Maxwell~\cite{Maxwell}. 
Structural rigidity concepts were first introduced in the study of
granular media by Guyon et al~\cite{Guyon}, who stressed that granular contact
networks differ from linear elastic networks in an important aspect: the first are
only able to sustain compressive forces between grains. 
Technically speaking, force networks with a sign-constraint on 
stresses are called \emph{struts}. Another typical example of sign-constrained networks
are spider webs, or cable structures, the elements of which (strings) can only
sustain traction forces~\cite{Tension}. Structures with interesting properties can be
obtained by combining elements of both types, in which case the resulting
framework is called a \emph{tensegrity structure}~\cite{tensegrity}. 
\begin{figure}[htb] 
\hbox{ \hsize=0.5 \hsize
{\bf \large a} \centerline{\psfig{figure=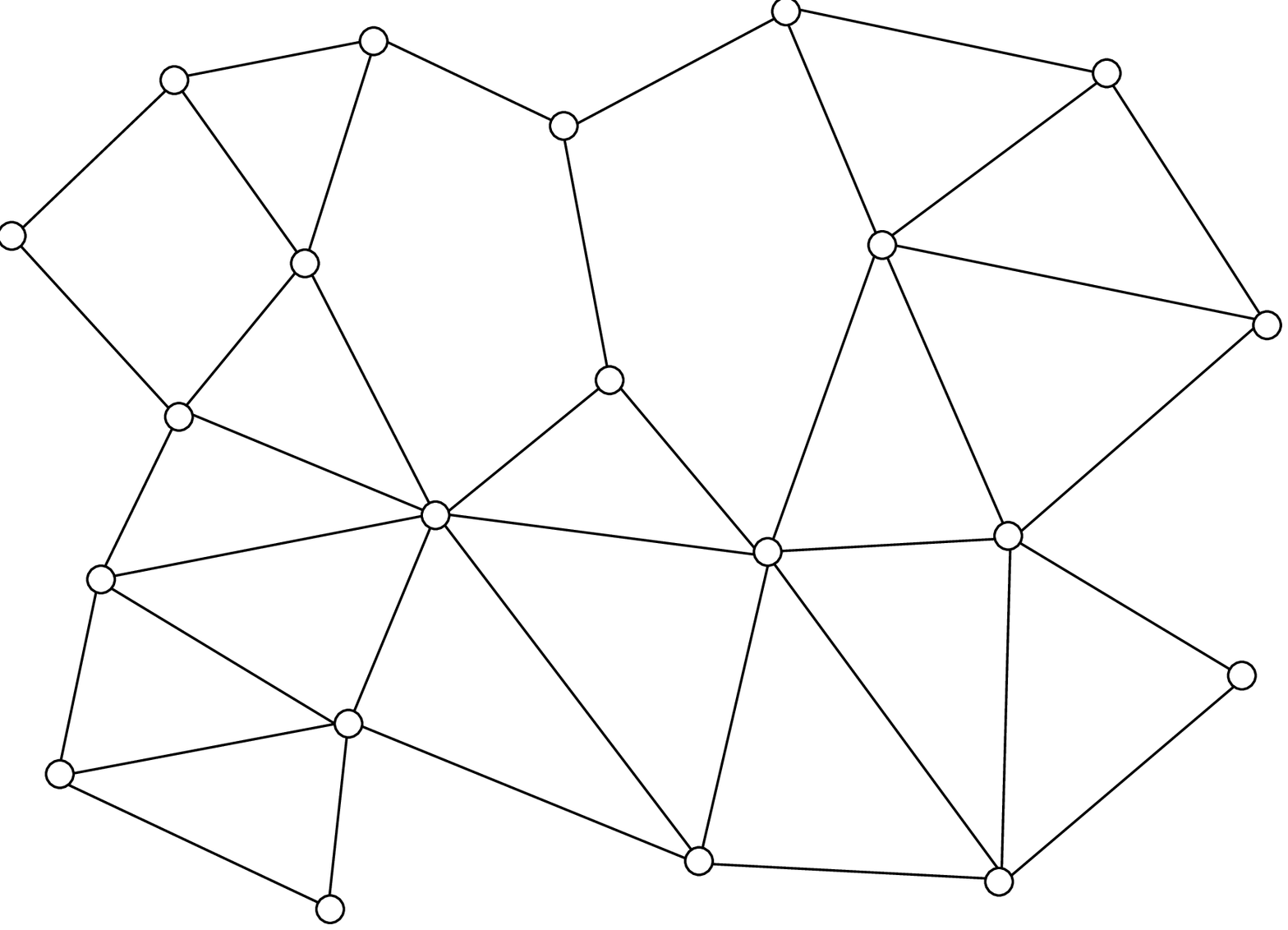,width=6cm,angle=270}}
{\bf \large b} \centerline{\psfig{figure=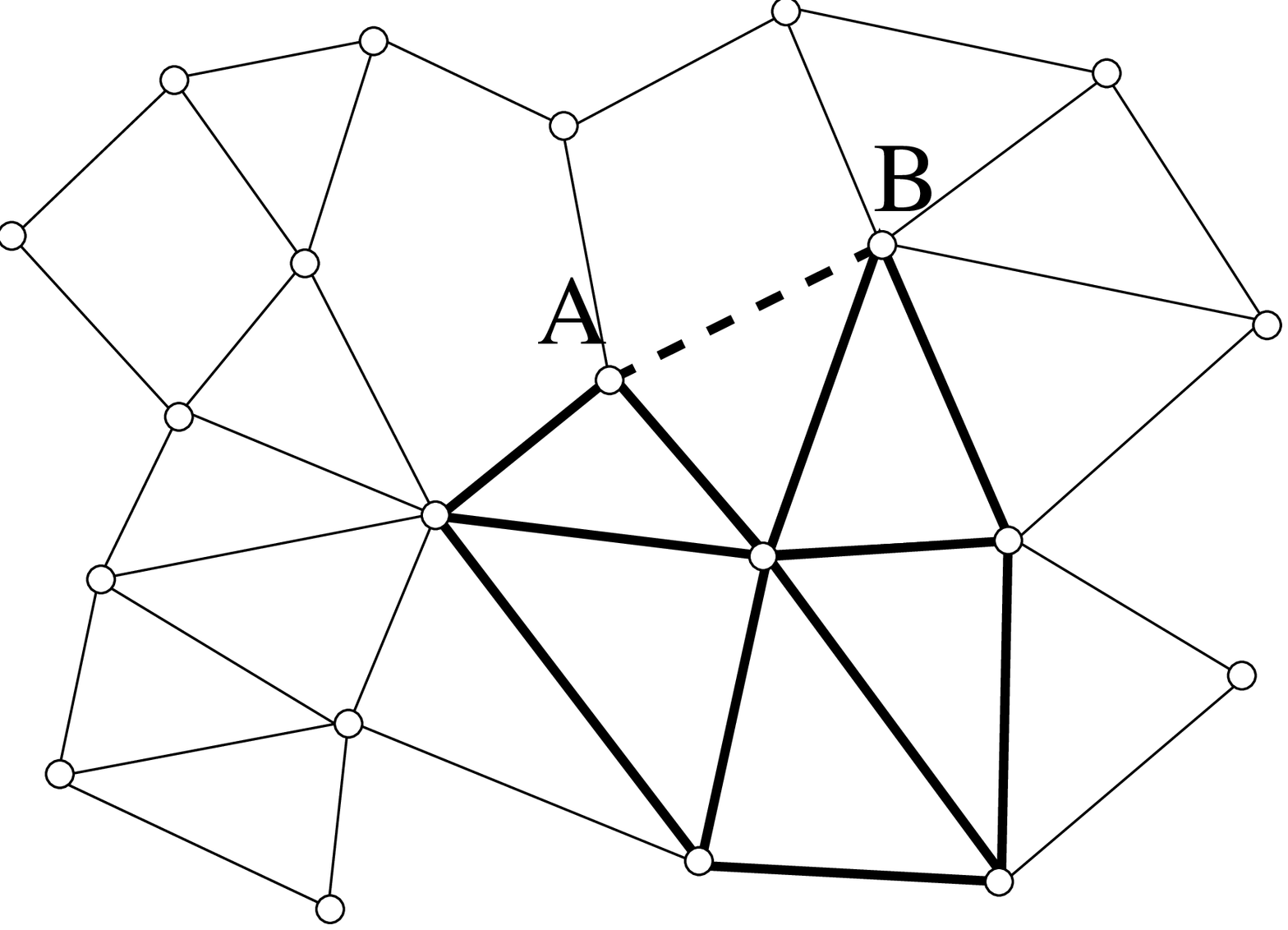,width=6cm,angle=270}}
}
\caption{ {\bf a)} A minimally rigid network in two dimensional space, composed of $n$
points and $2n-3$ bars. If any bar in this network is removed, some of
the points would cease to be rigidly connected to the rest. All bars in this
network are therefore essential for rigidity. {\bf b)} A 
\emph{redundant} bar (dashed line) between points A and B can induce
stresses (``self-stresses'') in some part of the network (dark lines). The
locus of self-stresses is the 
\emph{overconstrained} subgraph associated with bar AB. This is
exactly the set of bars which already provides a rigid connection between
points A  and B. If the length of bar AB is exactly equal to the distance
d(A,B), i.e. if there is no \emph{length-mismatch}, self-stresses will be
zero.} 
\label{fig:networks} 
\end{figure} 
Several applications of rigidity related ideas and
tools have been already presented in this book. Let us here only briefly refresh
some concepts which we need for our discussion. 

A point in $d$ dimensions has $d$ degrees of freedom, while a
rigid cluster has $d(d+1)/2$. Therefore if a set of $n$ points forms a rigid
cluster, it must be connected by at least $dn-d(d+1)/2$ bars. If a rigid
cluster of $n$ points has exactly $dn-d(d+1)/2$ bars, it is said to be
(generically) \emph{isostatic}, or minimally rigid (Fig.~\ref{fig:networks}a). 
A framework with more bars than necessary to be rigid is \emph{hyperstatic} or
\emph{overconstrained} (Fig.~\ref{fig:networks}b). 
Excess bars, which can be removed without introducing new degrees of freedom, are
called \emph{redundant}. A bar is redundant when the two points it connects are
already rigidly connected. Unless the length of the redundant bar is exactly
equal to the distance between the points it connects, self-stresses will be generated
in some parts of the framework. Thus self-stresses are non-zero only within
overconstrained subgraphs, and can be absent if there are no
length-mismatches.

We will discuss granular piles under the action of gravity, in which forces 
are transmitted to a supporting substrate. We could as well consider any other
load condition, provided that an infinitesimal gravity field is added in order
to remove indeterminacies in the positions of the particles. Because of the
action of gravity, contact networks associated to static granular piles must
be rigid. Otherwise some of the grains would be set in motion by gravitational
forces. Because of the fact that grains are rigidly connected to a  lower
boundary, a redundant contact anywhere on the system will usually produce an
overconstrained subgraph that extends all the way down to the rigid boundary.\\  
\begin{figure}[htb] 
\hbox{ \hsize=0.5 \hsize
\centerline{\psfig{figure=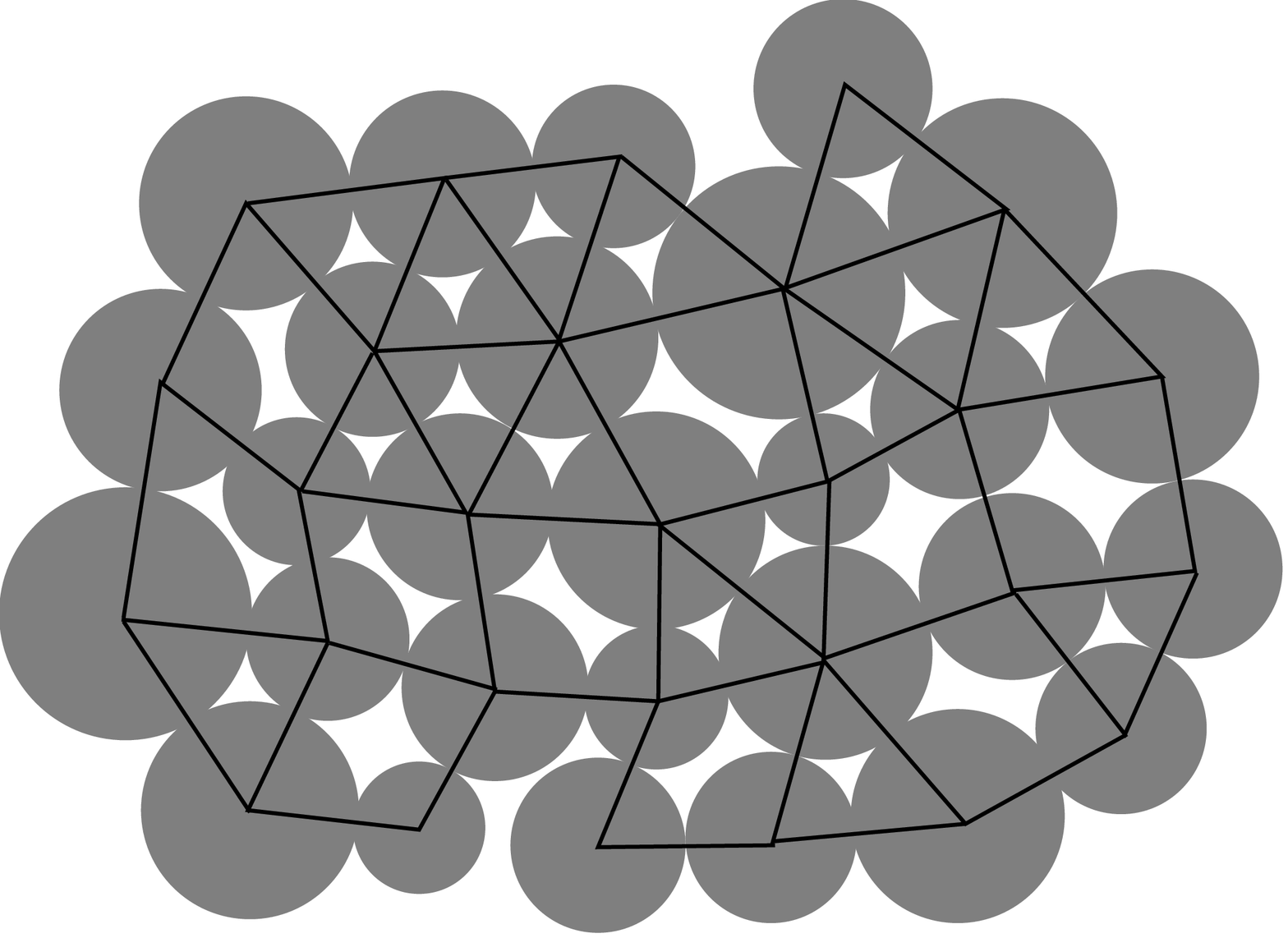,width=6cm,angle=270}}
\centerline{\psfig{figure=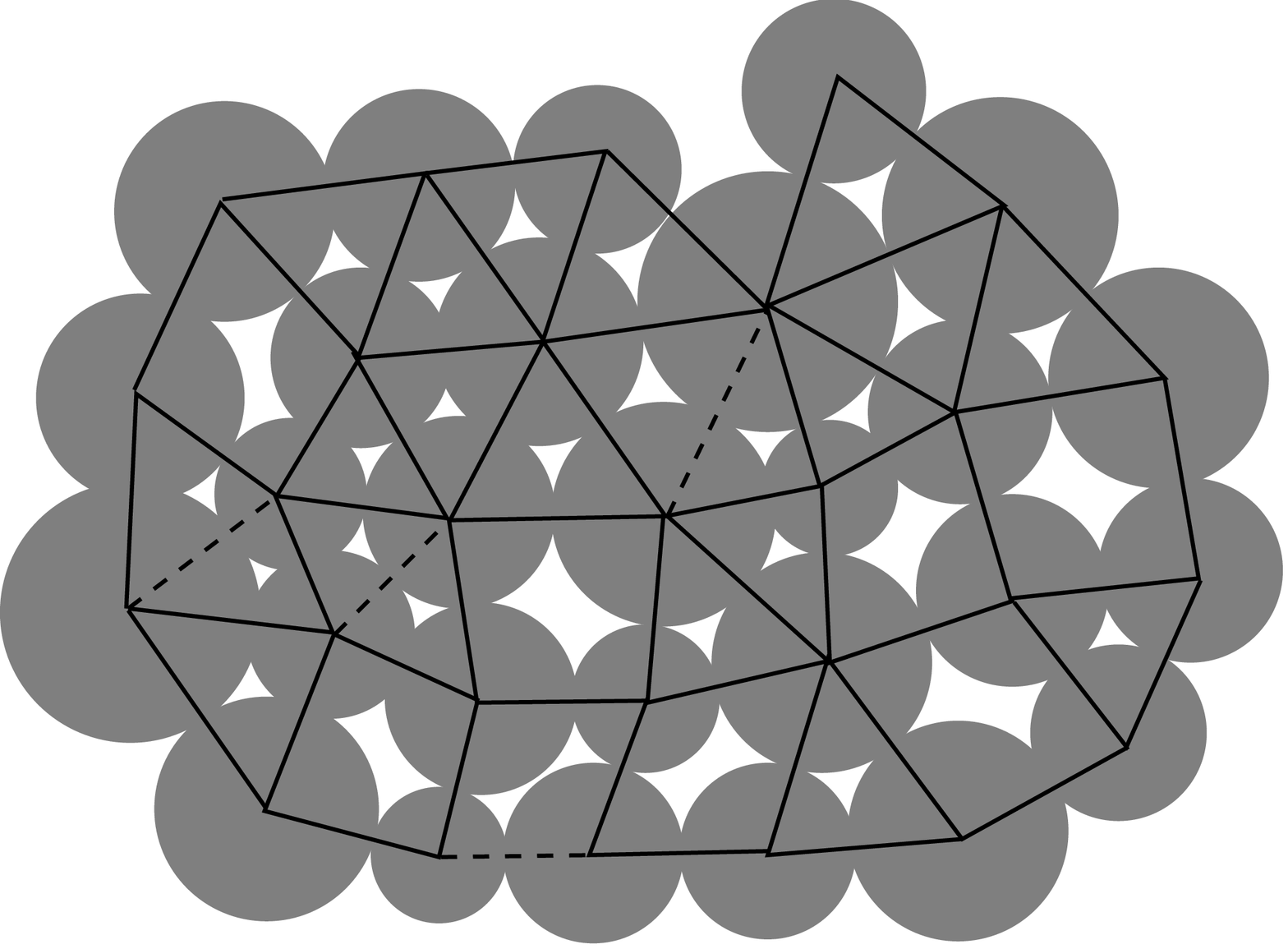,width=6cm,angle=270}}
}
\caption{ In the limit of large stiffness-to-load ratio, i.e. when the
compressive forces are small, or the rigidity large (left), the contact
network of a granular packing is sparse and, as discussed in the text, becomes
{\it minimally rigid}. If the compressive forces are increased, or the
stiffness decreased (right), excess contacts (dashed lines) are created. }
\label{fig:packs} 
\end{figure} 
Some early attempts to numerically study the static properties of granular
materials have ignored the sign constraint, thus modeling granular piles as
randomly diluted linear frameworks~\cite{Feng}. Due to this, it has been
sometimes suggested that rigidity percolation concepts might be applied to
granular networks~\cite{Guyon,RSH-G}. But this would require forces with
power-law distribution, since the elastic percolation backbone is a fractal
object at  the critical point~\cite{Feng,Letter}, while experimental and
theoretical studies~\cite{Liu,Miller,Radjai} suggest that force distributions
display exponential decay on granular systems.   

This suggests that the sign-constraint associated with non-cohesive granular
matter cannot be neglected. As we shall soon show, it is possible to see from a
topological point of 
view that the sign-constraint has far-reaching consequences for the static
behavior of granular aggregates. We demonstrate next that this restriction forces
a stiff granular system to choose, among all possible equilibrated contact
networks, only those with the specific topological property of minimal
rigidity.   

\subsection{\bf Isostaticity of stiff networks with a sign constraint}
\label{sec:isostaticity}

Consider now a $d$-dimensional frictionless granular pile in equilibrium 
under the action of external forces $\vec F_i$ (gravitational, etc) on 
its particles. We represent the contact network of this pile by means
of a linear-elastic central-force network in which two sites are connected by
a bond if and only if there is a nonzero compression force between the two
corresponding particles in the pile (Fig.~\ref{fig:contact-network}). 

If the external compression acting on the pile 
is increased, particles will suffer a larger deformation, and therefore the
number of interparticle contacts will increase (see Fig.~\ref{fig:packs}).
Equivalently, if compression forces are weakened, or the stiffness of the
particles made larger, the number of interparticle contacts, and thus the
number of bonds on the equivalent contact network, will be reduced because
there are no cohesive forces between particles. But there is a lower limit for
the number of remaining contacts, given by the condition that all particles be
rigidly connected, otherwise they would move until new contacts are
established. Therefore one may expect that in the limit of infinite
stiffness the resulting contact network will be minimally rigid. Let us now
formalize this observation.

Because of linearity, stresses $\fij$ on the bonds of the linear-elastic equivalent
network can be uniquely decomposed as 
\be
\fij=\fsij+\flij
\label{eq:superposition}
\ee \noindent
, where $\fsij$ are self-stresses, and $\flij$ are load-dependent stresses.  

Load-dependent stresses are linear in the 
applied load: if all loads are rescaled, $\flij$ are rescaled by the same
factor. But $\flij$ are not changed if {\it all} elastic constants, or
\emph{stiffnesses}, are multiplied by a factor.  

Self stresses in turn do not depend on the applied load. They are in general linear
combinations of terms of the form $\kij \eij$, where $\kij$ is the stiffness
of bond $ij$ and $\eij$ its length-mismatch (See
Fig.~\ref{fig:networks}). The length mismatch $\eij$ of a bond is defined as the
difference between its repose length and its length in an equilibrium
configuration under zero external loads. \\
\begin{figure}[htb] 
\hbox{\hsize=0.5\hsize
{\bf \large a)}
\centerline{\psfig{figure=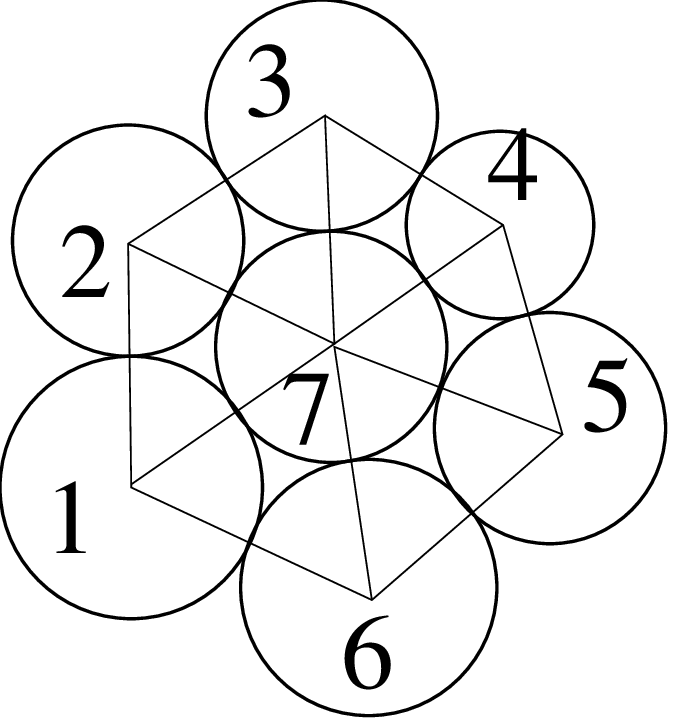,angle=270,width=5cm}}
{\bf \large b)}
\centerline{\psfig{figure=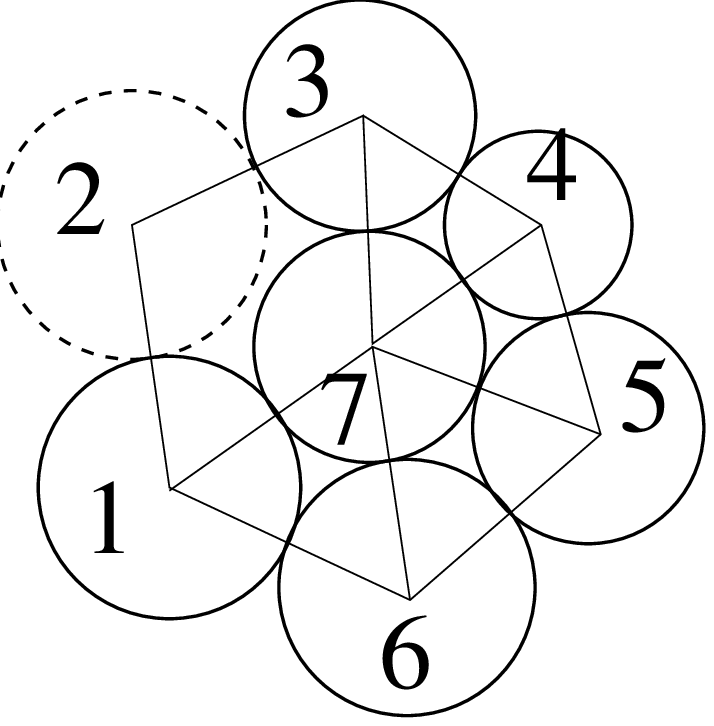,angle=270,width=5cm}}
}
\caption{{\bf a)} In order for seven particles to be in contact forming an
overconstrained graph as shown here, \emph{one} condition has to be satisfied
by the radii. One is free to choose the radii of 6 of them, but the seventh
will be uniquely determined by this choice.  {\bf b} If one of the particles
is for example slightly too large to satisfy the required condition, one
contact will be opened, restoring isostaticity. In this example the contact
between particles 2 and 7 is open, but any other bond between the central
particle and its neighbors could have been chosen. If on the other hand
particle 2 were too small, one of the contacts between the external particles
would be lost.
}
\label{fig:degenerate} 
\end{figure} 
As already discussed, length mismatches, and thus self-stresses, only arise
within overconstrained subgraphs~\cite{Crapo,Guyon}: those with more contacts,
or bonds, than strictly necessary to be rigid. For if a graph is not
overconstrained, then all its bonds, regarded as linear springs, can
\emph{simultaneously} attain their repose lengths while still being in
equilibrium under zero external load.  

It is easy to see that a bounded overconstrained subgraph with nonzero
self-stresses must have at least one bond subject to a negative (traction)
stress: As discussed, self-stresses are equilibrated without the intervention
of external forces. It then suffices to consider a joint belonging to the envelope
of the overconstrained subgraph. Since bonds can only reach it from one side 
of the frontier, self-stresses of \emph{both signs} must necessarily exist in
order for this joint to be equilibrated.

Now imagine rescaling all stiffnesses according to $k\to \lambda k$ (both in
the granular pile and in our equivalent elastic system). In doing so, all
self-stresses are rescaled by $\lambda$, but the load-dependent stresses
remain constant. Therefore, if self-stresses were non-zero, in the limit
$\lambda \to \infty$ at least one bond of the network would have negative
\emph{total} stress (eq.~(\ref{eq:superposition})). This is not possible since
traction forces are not allowed on our granular pile by hypothesis. 
Thus the existence of self-stresses is not possible in the limit of large
stiffness if there is a sign-constraint on total stresses. 

For reasons already reviewed, in order for self-stresses to be zero, one or
more of the following conditions will have to be satisfied by granular contact
networks in the limit of large stiffness: 
\vskip 0.2cm 
\noindent a) to have all length mismatches equal zero on overconstrained 
graphs.
\\
\noindent b) to have no overconstrained graphs at all (the network is isostatic). 
\vskip 0.2cm
Condition a) requires that, even when overconstrained graphs exist, particle
radii satisfy certain conditions in order to exactly fit in the holes left
by their neighbors (see Fig.~\ref{fig:degenerate}). But as soon as particles 
have imperfections or polydispersity (no matter how small if their rigidity is
large enough) self-stresses would appear if the contact network is hyperstatic.
In other words condition a) cannot be \emph{generically}
satisfied, if by generic we understand for a ``randomly chosen'' set 
of radii. Therefore condition b) must generically hold, i.e. there will be no
overconstrained subgraphs in the limit of large stiffness-to-load ratio.

From the point of view of this work, most experimentally realizable
packings fall under the category ``generic'', since small imperfections
in radius are unavoidable in practice. 

\vskip 0.3cm
\vbox{\noindent
In view of the above, we can now conclude that:
\begin{quote}the contact network of a  polydisperse granular pile becomes 
\emph{isostatic} when the stiffness is so large that the typical self-stress, 
which is of order $k\epsilon$ would be much larger than the typical
load-induced stress.
\end{quote}
}

Exceptions to this rule are packings with periodic boundary conditions,
because they are not bounded, and packings in which the radii satisfy exact
conditions in order to have zero length-mismatches in overconstrained graphs,
because they have no self-stresses~\cite{exceptions}. One can for example
consider a regular triangular packing of exactly monodisperse particles, in
which case the associated  contact network will be the full triangular
lattice~\cite{Oron}, i.e. hyperstatic. But the contact network (and the 
properties of the system as we shall soon see) will be drastically modified as
soon as a slight polydispersity is present~\cite{RSH-M,Luding,Travers} if the
stiffness is large enough. 
Therefore, while one is free to consider specific packings which are
not isostatic, in practice these cannot be realized for hard  particles,
since any amount of polydispersity, no matter how small, will force some
contacts to be opened so as to have an isostatic contact network. We now
discuss how this affects the properties of granular systems.

\section{CONSEQUENCES OF ISOSTATICITY}
\label{sec:consequences}

Isostaticity has been sometimes imposed in numerical
models~\cite{Jan}, as a condition allowing one to calculate 
stresses by simple propagation of forces. Recently isostaticity 
was reported by Ouaguenouni and Roux~\cite{OR}, who use an
iterative numerical algorithm to find the stable contact network of a set of
rigid disks. Our discussion in the previous section shows that isostaticity is
a generic property of stiff packings, and appears because negative stresses are
forbidden (an equivalent conclusion would be reached if only traction stresses
were allowed~\cite{Tension}). We will now show that isostaticity has important 
consequences on the way the system reacts when it is perturbed, but before
starting a more rigorous analysis, let us first discuss some of the most
important differences between isostatic and hyperstatic systems, on an
intuitive level.  

Imagine perturbing an elastic network by letting
an equilibrated pair of collinear forces act on the ends of a given bond, and consider
how \emph{stresses} and \emph{equilibrium positions} are modified in the whole
system. A properly chosen change in the repose length of this bond would have
exactly the same effect, so we can alternatively consider the
perturbation to be a change in length or a couple of forces.  
\begin{itemize}
\item
On overconstrained rigid
networks, stresses are \emph{correlated} over long distances: if a bond is
stretched as described above, stresses will be modified on other bonds far away from the
perturbation. This is so  because self-stresses \emph{percolate} through the
system. 

But for the same reason the \emph{displacements} of the sites from their
original positions, induced by this local perturbation, decay very fast with
distance. The reason for this is that self-stresses oppose the perturbation
and thus tend to ``quench'' its effect.  
\item
On isostatic systems, \emph{stresses are uncorrelated}. If we change the length of
(or apply a equilibrated pair of forces to) an arbitrary bond, stresses on all
other bonds remain unchanged, because they only depend on external loads,
and not on bond lengths. This is a trivial property of isostatic systems.

But the equilibrium position of many sites will be in general modified if one
of the lengths is changed, and therefore displacements induced by a
perturbation may be felt far away from its origin. \\
\end{itemize}
\begin{figure}[htb]
\centerline{ \psfig{figure=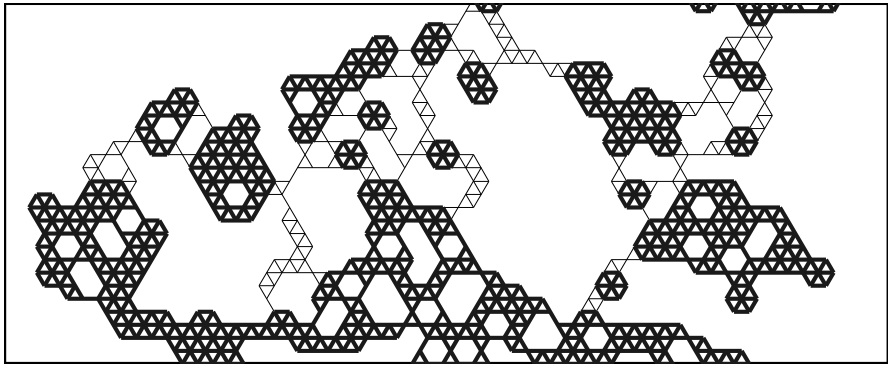,width=12cm}}
\caption{Part of the stress-carrying backbone for rigidity percolation on a 
randomly diluted triangular lattice. Overconstrained rigid clusters (thick
lines) are isostatically connected to each other by \emph{cutting bonds} (thin
lines). Cutting bonds are critical in the sense that they provide a minimally
rigid connection. All of them are essential for rigidity. }
\label{fig:backbone}
\end{figure}

Therefore on a hyperstatic system, perturbations of stresses propagate over
long distances,  while on isostatic systems it is the displacement field which
may display long-range correlations. Stresses are uncorrelated on isostatic
systems, and thus arbitrarily large stress gradients are possible. Hyperstatic
systems, on the opposite hand, have smoothly varying stresses because of the
strong correlations introduced by overconstraints. This provides some
indication that stress concentration is possible \emph{because} of
isostaticity. \\
We can get further insight into the meaning and potential consequences of
isostaticity from recent studies of central-force rigidity
percolation~\cite{Letter}. Rigid backbones, the stress-carrying components of
rigidly connected clusters, 
are found to be composed of large overconstrained clusters,
\emph{isostatically} connected to each other by \emph{critical bonds} (also
called red bonds, or cutting bonds -- See Fig.~\ref{fig:backbone}). 
Overconstrained clusters have more bonds
than necessary to be rigid, so any one of them can be removed without
compromising the stability of the system. But the rigid connection among these
clusters, provided by critical bonds, is isostatic, or minimally rigid. 
In other words, cutting one critical bond is enough to produce the
collapse of the entire system, because each isostatic bond is by definition
\emph{essential} for rigidity. Thus we may expect that stretching a critical
bond will have a measurable effect on a large number of sites.

But in percolation backbones, critical bonds only exist very close to the
rigidity percolation density $p_c$. Above $p_c$ there is percolation of 
self-stresses~\cite{Letter} and thus the rigid backbone is hyperstatic. Even
exactly at $p_c$ the number of critical bonds is not extensive, but scales as
$L^{1/\nu}$ where $\nu$ is the correlation-length
exponent~\cite{Letter,Coniglio}. Consequently critical bonds are relatively few at
$p_c$, and virtually absent far from $p_c$. Thus if we perturb (cut or
stretch) a randomly chosen bond in a percolation backbone, most of the times
the effect will be only be local since no critical bond will be hit. 

The important new element in stiff granular contact-networks is the fact that
\emph{all} contacts are isostatic, i.e. there is \emph{extended isostaticity}.
In this case, if any of the bonds (contacts) where removed, the pile would cease to be
rigidly connected to the supporting boundary below it. Because of this, if the
length of \emph{any} of the network's bonds is changed (which corresponds to a
variation in one of the particles radii) the equilibrium position of a finite 
fraction of the particles will also be changed. For these reasons, one may
expect that isostaticity will produce a large sensitivity to perturbation in
granular networks.

\subsection{\bf Susceptibility to perturbation}
\label{sec:susceptibility}

We now quantify the degree of susceptibility to perturbation, and then see
whether the intuitively appealing ideas we have just discussed are in fact 
verified on specific models. In order to provide a formal definition of
susceptibility, we introduce an infinitesimal change in the length $l_{ij}$ of
a randomly chosen bond of the network, and  record the \emph{induced
displacement} $\vec \delta_i$ suffered by all particle centers. We then define
the system's susceptibility $D$ as 
\be
D=\sum_{i=1}^N {\vec \delta_i}^2
\ee \noindent
, where $N$ is the total number of particles on the system. 
\begin{figure}[htb]
\centerline{\psfig{figure=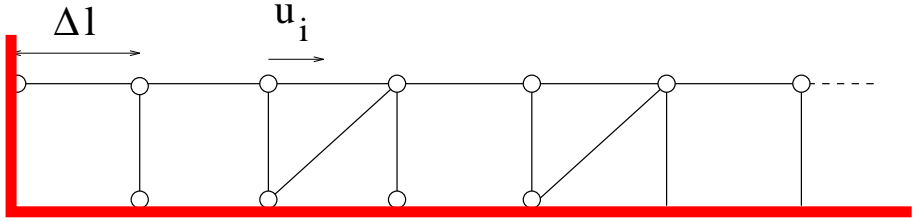,width=12cm,angle=270}}
\vskip 0.2cm
\caption{ A linear chain of springs in which overconstraints
(diagonals) are present with probability $O_v$ is useful as a toy model to
understand the influence of isostaticity on the propagation of perturbation.
The characteristic distance for the decay of displacements induced by a 
perturbation (a bond-stretching) diverges in the
limit $O_v \to 0$ and thus the susceptibility to perturbation also
diverges (see text). }
\label{fig:onedim} 
\end{figure}
These measurements are done for variable amounts of \emph{overconstraints}
(excess contacts) 
randomly located on an elastic network, and averages are performed over disorder. 
In this way $D(O_v)$ is obtained, where $O_v$ is the \emph{density of
overconstraints}.  
According to our previous discussions, we expect $D$ to increase 
as $O_v \to 0$, which is the isostatic limit. 

We start by discussing a toy model shown in Fig.~\ref{fig:onedim}: a quasi
one-dimensional system composed of linear elastic bonds, in which
diagonals are present with probability $O_v$. The system with no diagonals
($O_v=0$) is exactly isostatic, therefore each diagonal is an overconstraint,
or redundant bond. We assume for simplicity that all bonds have the same
stiffness $k$ and length $l$ (diagonals have a length $\sqrt{2}l$). We are
interested in calculating the average horizontal displacements $u(x)$ induced
by a length change $\Delta l$ in the left-most horizontal bond, as a function
of the density $O_v$ of overconstraints (diagonals). A simple
calculation~\cite{tbp} shows that, after averaging over disorder, the
displacement field $u(x)$ satisfies
\be
\frac{\partial^2 u}{\partial x^2} = \kappa ^2 O_v u
\ee \noindent
, where $\kappa$ is some constant.   
Therefore \hbox{$u(x)=u(0) \exp\{-\kappa O_v^{1/2} x\}$} and we see that
there is exponential decay with distance, with a characteristic
length $\xi(O_v) \sim O_v^{-1/2}$. This ``persistence length'' diverges at
$O_v=0$, which corresponds to the isostatic limit. 
Consequently $O_v=0$ is a critical point, and $D$ as defined above is
divergent there. The divergence of $D$ in this model is linear with system
size. \\
\begin{figure}[htb]
\vbox{
\centerline{{\bf a)} \psfig{figure=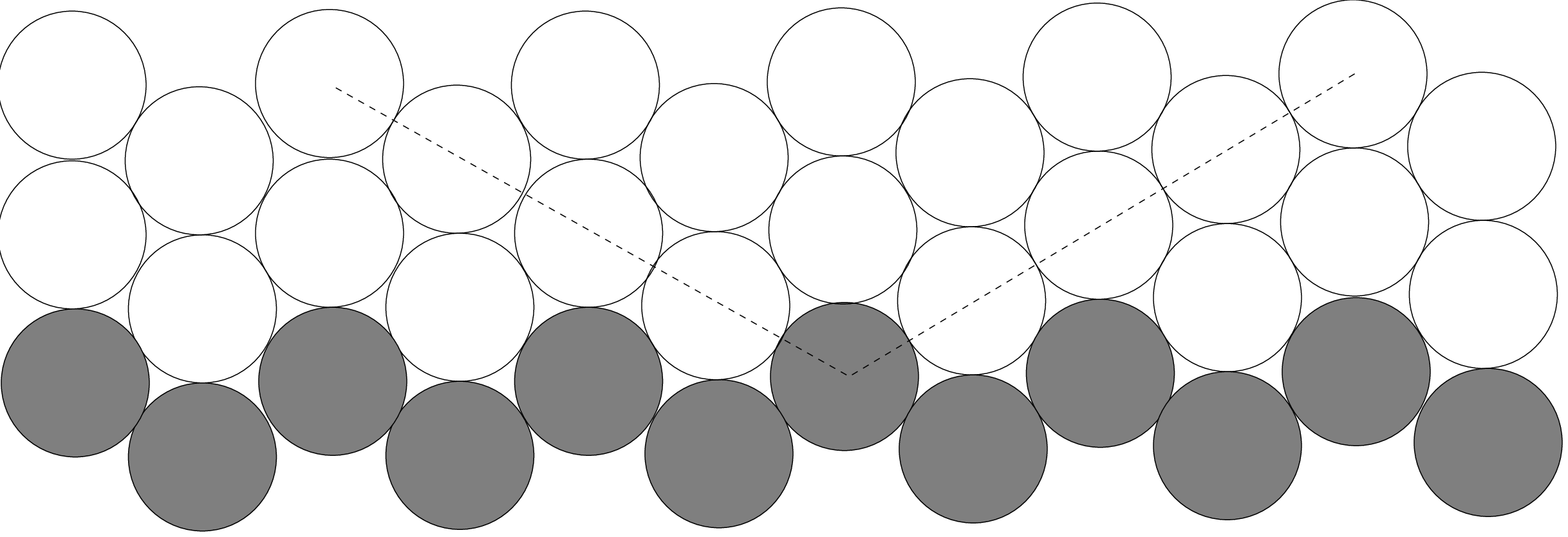,width=12cm,angle=270}}
\vskip 0.3cm
\centerline{{\bf b)} \psfig{figure=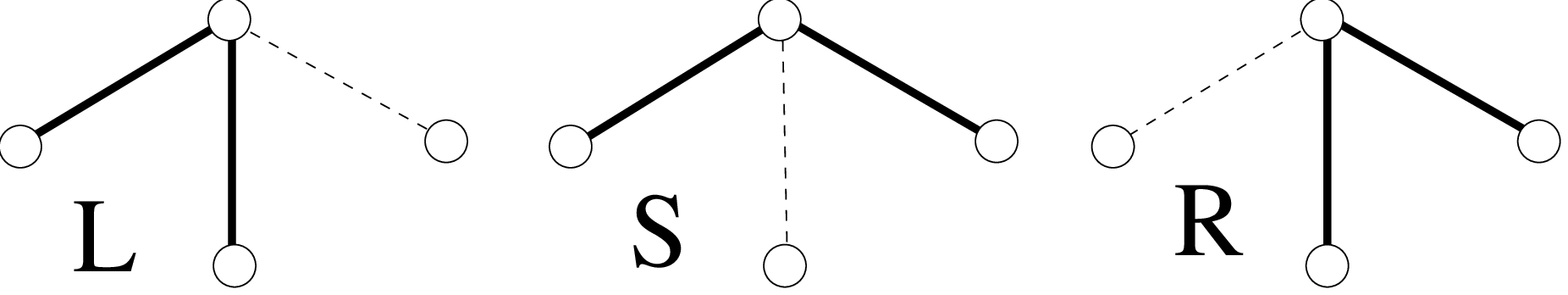,width=12cm,angle=270}}
}
\caption{
{\bf a)} In two dimensions, a triangular packing is used in order to
numerically measure susceptibility to perturbation. The two first layers of
particles (shaded) is regarded as a fixed rigid boundary supporting the load
of the upper ones. If a site is shifted, only particles within its ``cone of
influence'' (dashed line) can be displaced. In this example, 6 layers of 6
particles each are displayed.
{\bf b)} Appropriately choosing among these three isostatic configurations 
for each site, only compressive stresses are produced. First S
is chosen with probability 1/2. If S is not chosen then either R or L are, 
depending on the sign of the horizontal force acting on the site (see text).
}
\label{fig:twodim} 
\end{figure}
Now let us see whether isostaticity has comparable effects in two
dimensions. In the spirit of
previously proposed models~\cite{RSH-G,Jan} we consider a triangular packing
oriented as in Fig.\ref{fig:twodim}a, made of very stiff disks with small
polydispersity, under the action of gravitational forces. The polydispersity
is assumed to be small enough such that disk centers are approximately located
on the sites of a regular triangular lattice, and the stiffness to load ratio
large enough such that the contact network is isostatic, according to our
discussion in the previous section. 

The full problem of generating realistic contact networks that respect the
constraint of no traction forces is a difficult one. Several approaches
have been proposed~\cite{RSH-G,Jan}, all involving some degree of approximation
even for geometrically simple settings. In all these models, the triangular 
lattice was oriented with one of its principal axis horizontal, i.e.
normal to gravity. There is though some advantage in considering a
different orientation, such that one of the principal axis of the 
lattice is parallel to gravity (Fig.~\ref{fig:twodim}). In this
case there is no need for recursive checks of positiveness
of stresses since the disordered contact network can be built in a fashion
that guarantees positive stresses. Our model is defined as follows:
We ask that each site be supported by exactly two out of its three
lower neighbors, thereby ensuring that only isostatic contact networks
are generated. This condition gives three possible local configurations which
we call left (L) symmetric (S) and right (R) respectively and are depicted in
Fig.\ref{fig:twodim}b. Choosing a local configuration of bonds on each site
produces a sample, or one realization of the disorder. 
Clearly not any choice of bond configurations give rise to a contact network
with positive stresses. But it is possible to satisfy the
positivity constraint and still have disorder in the  following way:
\begin{enumerate}
\item For each site, starting from the uppermost 
layers and proceeding downwards, we choose configuration S with probability 
$1/2$~\cite{symmetric}. 

\item If S was not chosen, then either R or L are, according to the sign of the
horizontal force-component $F_x$ acting on that particle:  
if $F_x >0$ ($F$ points rightwards), R is chosen. If $F_x<0$,
then L is chosen. If the horizontal component is zero then R or L are chosen at random.
\end{enumerate}

Our model has no geometrical disorder, and this is justified by our assumption
of small polydispersity, but we keep contact disorder and isostaticity which
are the important characteristics of real granular networks in the limit of
large stiffness.  

There is no reason to think that the method we have chosen generates all
possible 
equilibrated contact networks that satisfy isostaticity and positiveness of
stresses.  It seems in principle possible to have some sites making contact to
all three downward neighbors and still have isostaticity, by simultaneously
opening some other contacts. But our aim here is not to provide a realistic
model for granular contact networks but to test whether isostaticity has
important effects on the properties of a two-dimensional network.

In order to accomplish this we will measure, on the networks so generated, the
susceptibility defined above, and compare the results with those obtained
on systems with a finite density $O_v$ of overconstraints randomly located
on the network. A non-zero density of overconstraints $O_v$ mimics, as
discussed previously, the effect of increasing the mean pressure on the system
(or reducing the stiffness), since this would produce a larger number of
contacts, in excess of isostaticity, to be established between particles. 

A finite density $O_v$ of overconstraints is introduced in this model by
letting \emph{all three bonds} be connected below a given  site, with
probability $O_v$. Each third bond  introduced in this way creates an
overconstrained subgraph that extends all the way down to the rigid
boundary. The limit $O_v=1$ gives the fully connected triangular lattice,
which of course has no disorder. 
\begin{figure}[htb]
\hbox{\hsize=0.5\hsize
{\bf \Large  a}\centerline{\psfig{figure=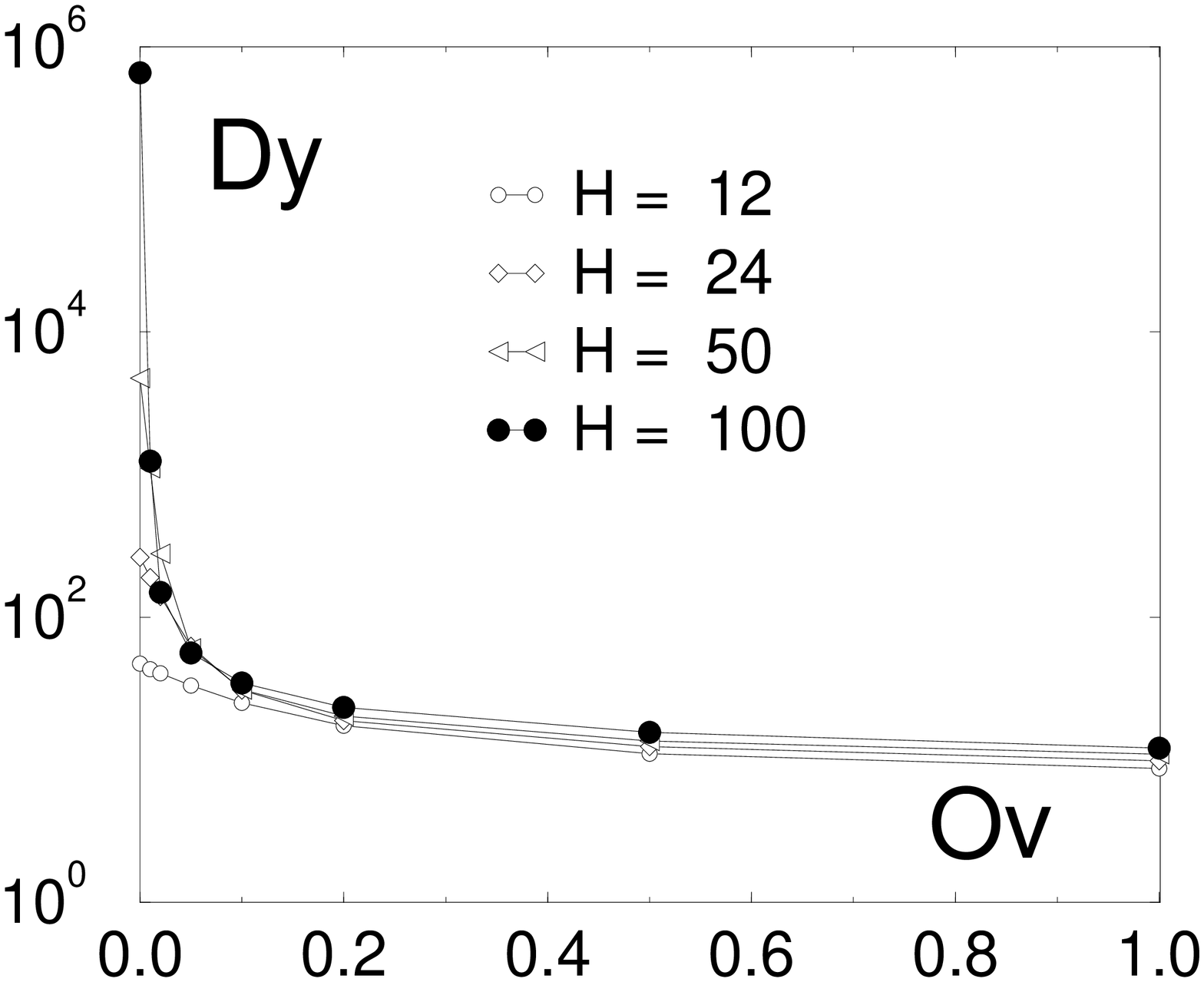,width=6cm,angle=270}}
{\bf \Large  b}\centerline{\psfig{figure=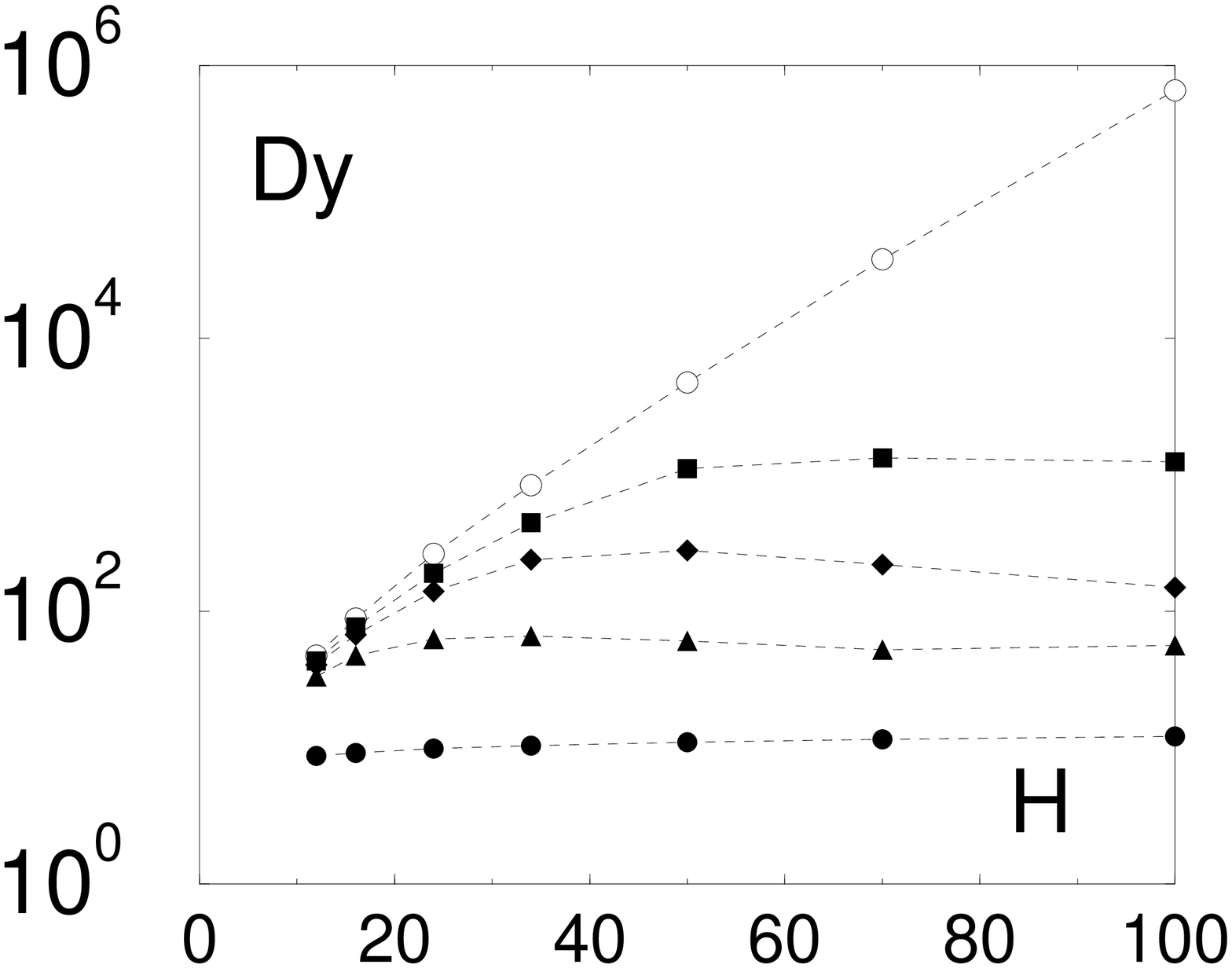,width=6cm,angle=270}}
}
\caption{
{\bf a)} Susceptibility $D_y$ (defined in the text), versus density of
overconstraints $O_v$, as measured on two-dimensional triangular packings of
total height $H=12, 24, 50$ and $100$ layers. $O_v=0$ is the isostatic limit,
and corresponds to granular packings with infinite stiffness, as demonstrated
in the text.  
{\bf b)} Susceptibility $D_y$ versus system height $H$, for fixed fractions
$O_v$ of overconstraints: 0.00 (empty circles), 0.01 (squares), 0.02
(diamonds), 0.05 (triangles), 1.00 (full circles). It is clear in this figure
that, for $O_v>0$ , $D_y$ {\it
saturates} to a finite value in the $H\to \infty$ limit. For isostatic
systems, on the other hand, $D_y$ diverges {\it exponentially fast} with $H$.
See also Fig~\ref{fig:results_cd}a).
}
\label{fig:results_ab}
\end{figure}
After building a contact network with a specified density of overconstraints
as described above, an infinitesimal upwards shift is introduced
in a randomly chosen site on the lowest layer, and the induced
displacement field $\vec \delta_i$ is measured. 

If the network is isostatic ($O_v=0$) one can
calculate all stresses~\cite{RSH-G,Jan} and displacements~\cite{tbp} in
a \emph{numerically exact fashion} so that systems of $2000\times 2000$ 
particles may be simulated on a workstation. The idea is that stresses
are propagated downwards and displacements upwards. The way in which the
induced displacements are propagated upwards is easily calculated~\cite{tbp}
by noting that, when the network is isostatic, all bond lengths (except the
perturbed one) must remain \emph{constant}. 
On the other hand when the network is overconstrained ($O_v > 0$ ), stresses
and displacements can no longer be exactly calculated. In this case one has to
solve the elastic 
equations in order to find the new equilibrium positions after the perturbation. 
This is done in the limit of linear elasticity (since the
perturbation is infinitesimally small) by means of a conjugate gradient
solver. In this case the calculations are much more time-consuming so that
only relatively small systems, or order $100 \times 100$ particles can be
studied if $O_v \neq 0$. Supercomputers are required for this part of the
calculation~\cite{HLRZ}. A cross-check of the computer programs was done by
comparing the results obtained with the direct solver for isostatic systems,
with those produced by the conjugate gradient solver with no
overconstraints, on systems of up to $100\times100$ particles. Excellent
agreement was found in all cases, for stresses as well as as for displacements.   

In this way the susceptibility $D_y(H,O_v) = <\sum_{i=1}^N {\delta y_i}^2>$ 
is measured, where $\delta y_i$ is the vertical displacement of site $i$ due to the
perturbation, and $<>$ stands for average over disorder realizations. The
system consists of $H$ layers of $H$ grains each, so that $N=H\times H$.
\begin{figure}[htb]
\hbox{\hsize=0.5\hsize
{\bf \Large a} \centerline{\psfig{figure=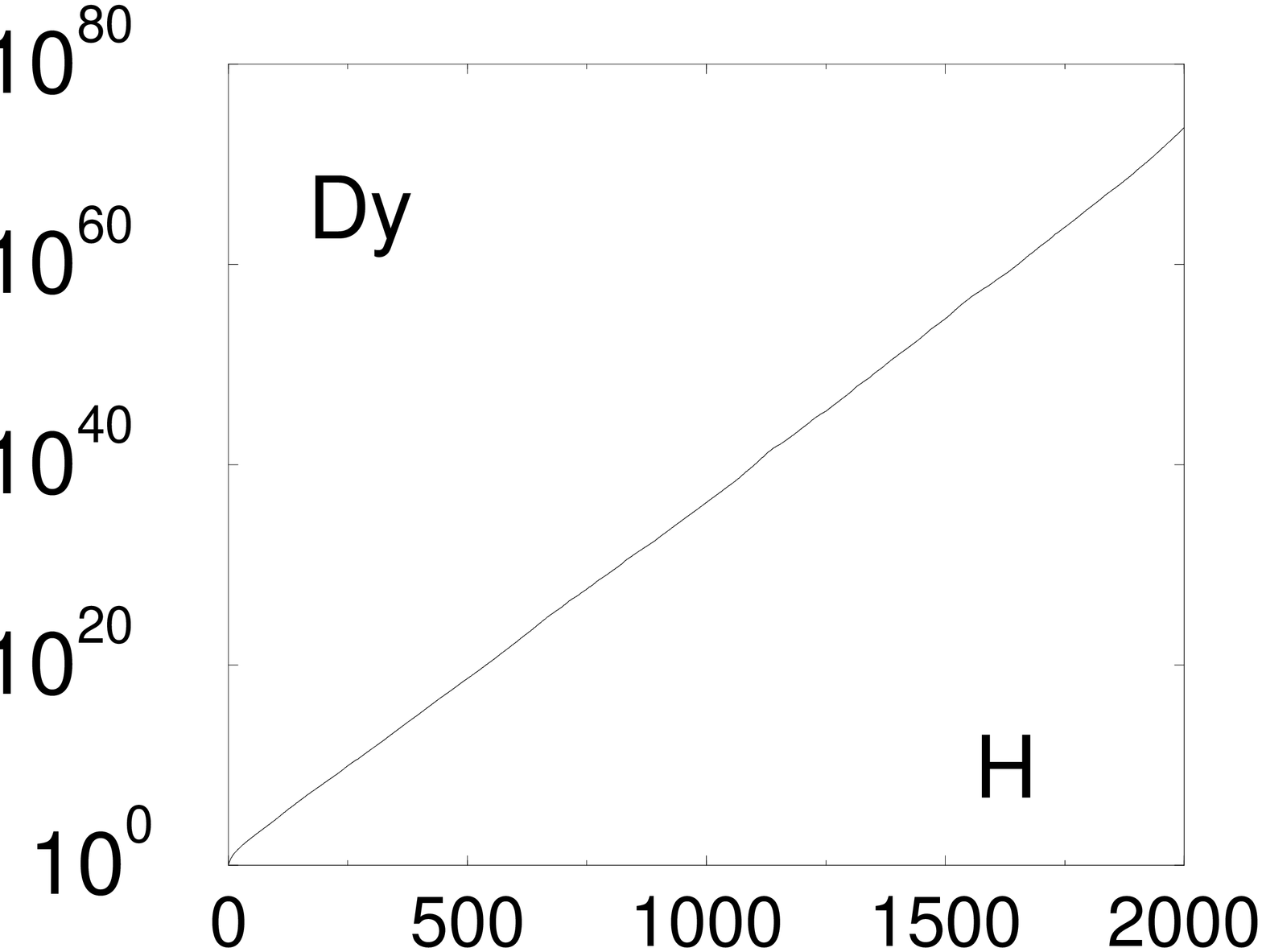,width=6cm,angle=270}}
{\bf \Large b} \centerline{\psfig{figure=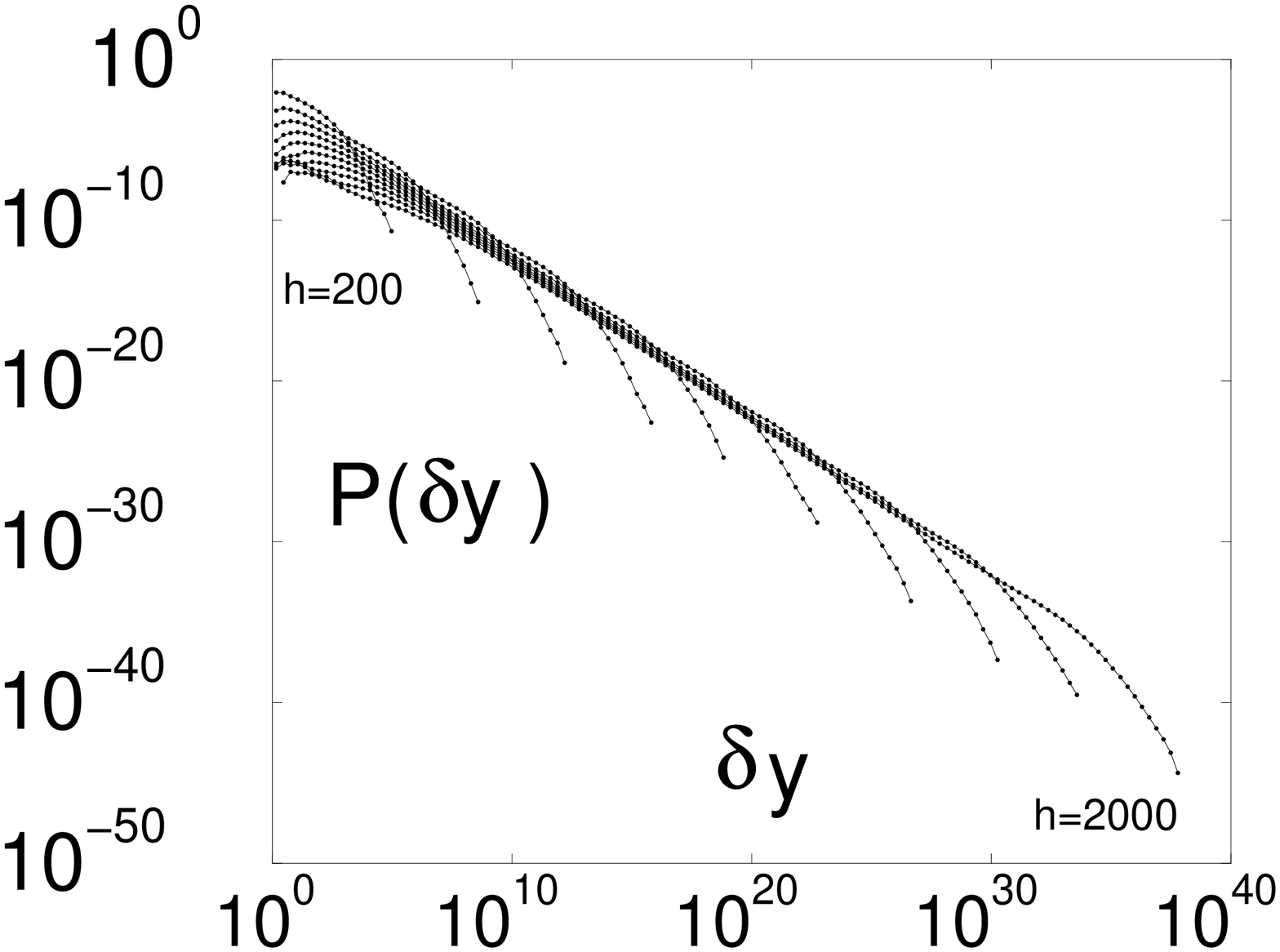,width=6cm,angle=270} }
}
\caption{
{\bf a)} $D_y$ grows
\emph{exponentially} with height when $O_v=0$. {\bf b)} The probability
$P_h(\delta y)$ to have an induced displacement $\delta y$ is power-law
distributed on isostatic networks. Results are shown at 
$h=200,400,600,\ldots,2000$ layers above the perturbation, for positive
values of $\delta y$ only. $P_h(\delta y)$ is approximately symmetric,
and has a finite peak of weight 0.47 at $\delta y=0$. Thus at any height above
the perturbation, approximately one half of the sites within the influence
cone are not vertically shifted.
}
\label{fig:results_cd}
\end{figure}
Figure \ref{fig:results_ab}a shows the susceptibility $D_y$ as a function of
the density 
of overconstraints $O_v$, for several system heights $H$. We see that $D_y$
increases rapidly on approach to the isostatic limit $O_v=0$, and that this 
increase is faster for larger systems, meaning that $D_y$ diverges at
$O_v=0$ in the $H \to \infty$ limit. Figure \ref{fig:results_ab}b shows the
same data now plotted as a function of system size, for several values of the
density of overconstraints. For any  $O_v \neq 0$, $D_y$ goes to a \emph{finite
limit} for large sizes, while it diverges with system size if $O_v=0$.
Data for much larger systems can be obtained in the isostatic case
using the direct solver program, and are displayed in
Fig.~\ref{fig:results_cd}a. This plot shows that $D_y(O_v=0)$ is of the form
$\log{D_y} \sim H $, that is, $D_y$ diverges \emph{exponentially fast} with
system size.   

These numerical results demonstrate that a \emph{phase transition} occurs at
$O_v=0$, where anomalously large susceptibility sets in. In a way which is
consistent with our intuitive expectations in the previous section, and 
with the one dimensional toy model, isostaticity is also in two
dimensions responsible for a large susceptibility to perturbation. An
important and surprising difference is the fact that, at the isostatic
critical point 
$O_v=0$, the susceptibility $D$ increases exponentially fast with system size,
whereas it grows only linearly with size in one dimension. 

Seeking to understand the surprisingly fast growth of $D$ with system size,
the probability distribution $P_h(\delta y)$ to have a vertical displacement
$\delta y$, $h$ layers above the perturbation has been measured on isostatic
systems of $2000 \times 2000$ particles. Only sites within a 120 degree cone
whose apex corresponds to the perturbed bond may feel the effect of the 
perturbation (Fig.~\ref{fig:twodim}a). $P_h(\delta y)$ thus gives the
probability for a randomly chosen site inside the influence cone of the
perturbed bond and $h$ layers above it, to have a vertical displacement
$\delta y$. Sites outside this cone have $\delta=0$, so our measurements
essentially correspond to a system of infinite width. 

Figure \ref{fig:results_cd}b shows the result of these measurements at
$h=200,400,600,\ldots,2000$ layers above the perturbation. Only positive
values of $\delta$ are displayed in this figure since $P_h(\delta y)$ is
approximately symmetric.  $P_h(\delta y)$ is found to be consistent with a
power-law behavior with an $h$-dependent cutoff:  
\be
P_h(\delta y) \sim h^{-\rho}|\delta y|^{-\theta} 
\ee
for $\delta y < \delta_M(h)$. 

It is also evident from Fig.~\ref{fig:results_cd}d, 
that the cutoff $\delta_M(h)$ grows exponentially with
increasing distance $h$ from the perturbation. Fitting the curve corresponding
to $h=2000$ in the interval $10^{10}<\delta y<10^{30}$, an estimate
$\theta=0.98$ is obtained, suggesting that $\theta \to 1$
asymptotically. Normalization then requires $\rho=1$, since the cutoff
$\delta_M(h)$ increases exponentially with $h$.  

Similar measurements where done for (smaller) systems with a finite density of
overconstraints $O_v$, in which case the distribution of displacements
presents a size-independent bound (Fig.~\ref{fig:deltaov}b).

Thus, in two dimensions an isostatic phase transition takes place at $O_v=0$,
and the resulting isostatic phase is characterized by a susceptibility to 
perturbation that grows exponentially fast with system size. The distribution
of induced displacements is power law, with a cutoff that grows exponentially
with distance from the perturbation. 
Of course one does not expect to be able to really measure exponentially large
values of displacements on granular systems. The calculations reported here
are valid for infinitesimal displacements, and in with this in mind the contact
network is considered to remain unchanged during the perturbation. In
practice, internal rearrangements would occur before we could detect
very large values of displacement on a real pile. So how can
we know if the huge susceptibility to perturbation that our calculations
predict have any observable effect? This is is discussed in the next section. 

\subsection{\bf Isostaticity implies instability}
\label{sec:response}

In order to clarify the relevance of the findings described in
section~\ref{sec:susceptibility} in relation with the observed unstable
character of granular packings~\cite{Dantu,Travers,Liu,Miller}, we must
first demonstrate the equivalence between induced displacements on site $i$ and the
load-stress response function $\g_ib$ of the stretched bond $b$ with respect
to a load on site $i$.   

The network's total energy can be written as 
\be
E=\sum_{i=1}^N W_i y_i + 1/2 \sum_{b} k_b (l_b - l^0_b)^2
\ee \noindent
, where the first term is the potential
energy ($W_i$ are particle's weights) and the second one is a sum over all
bonds and accounts for the elastic energy.  
$l_b$ are the bond lengths in equilibrium and $l^0_b$ their repose
lengths (under zero force). 
Upon infinitesimally stretching bond $b'$, equilibrium requires that
\be 
\sum_i W_i \frac{\partial y_i}{\partial l_{b'}} + \sum_{ov} k_{ov} 
( l_{ov} - l_{ov}^0) \frac{\partial
l_{ov}}{\partial l_{b'}} = 0 
\ee 
where the second sum goes over bonds $ov$ that belong to the same
\emph{overconstrained} graph as $b'$ does. This is so since bonds not
overconstrained with respect to $b'$ \emph{do not change their lengths} 
as a result of stretching $b'$. Since stress $f_b$ on bond $b$ is
$f_b=k_b(l_b^0-l_b)$ this may be rewritten as
\be
 \sum_{ov} f_{ov} \frac{\partial l_{ov}}{\partial l_{b}} =
\sum_i W_i \frac{\partial y_i}{\partial l_{b}} 
\label{eq:response}
\ee
If there are no overconstrained graphs the left hand sum only contains bond
$b$ itself, therefore, 
\be
f_{b}= \sum_i W_i \frac{\partial y_i}{\partial
l_{b}}
\label{eq:sum}
\ee
showing that, in the isostatic case, the displacement 
\hbox{$\delta y_i^{(b)}=
\frac{\partial y_i}{\partial l_b}$} induced on site $i$ by a stretching of
bond $b$ is the \emph{response function} $\g_ib$ of
stress $f_b$ with respect to an overload on site $i$. 

Taking averages with respect to disorder on equation (\ref{eq:sum}), we obtain
\be
<f_b>= \sum_i W_i <\delta y_i^{(b)}>
\ee \noindent
, and since average stresses on a given
layer grow linearly with depth, we must have 
\be
<\delta y_i^{(b)}>_H  \sim H^{-1}
\label{eq:deltav}
\ee 
We have seen that the second moment of $P_H(\delta y)$ diverges as $\exp\{H\}$,
while (\ref{eq:deltav}) shows that its first moment goes to zero with increasing
$H$. This can only happen if  $P_H(\delta y)$ is approximately symmetric (this is
numerically verified), which demonstrates that large positive and
negative values of $\delta y$ appear with similar probability. 
Given now the equivalence between induced displacements and the load-stress
response function, the existence of large
negative induced displacements means that a  positive
overload at a random site $i$, would often produce a (very large) negative stress on
any arbitrarily chosen bond $b$. This in turn indicates that the system will
have to rearrange itself in order to restore compressive forces, since negative 
stresses are not possible. In other words, isostatic packings are \emph{unstable} to
small perturbations, and will reorganize themselves on the slightest change in
load, in order to find a new stable (compression only) contact network.

In order to finish the demonstration that {\it instability is a consequence of
isostaticity}, we still have to show that a finite density of overconstraints
would make the response function bounded again. 

When there are overconstraints, $P_H(\delta y)$ is no longer critical but
bounded as our numerical simulations show (see Fig.~\ref{fig:deltaov}). But
in this case $\delta y$ is no longer identically equal to the load-stress response
function, i.e. (\ref{eq:sum}) no longer holds. One can nevertheless see, by
looking at formula (\ref{eq:response}), that the weight-stress response
function $\g_ib$ must be bounded if $P_H(\delta y)$ is. 
Therefore in the overconstrained case a finite
overload of order $<f>$  is necessary in order to produce rearrangements, and
the system is thus no longer unstable.

\subsection{\bf Pantographs}
\label{sec:pantographs}

The exponential growth of $\delta_M(H)$ is  responsible for the observed
exponential behavior of the total susceptibility in the isostatic case. But we
have yet to understand for which reason exponentially large values of
displacements do exist.
\begin{figure}[htb]
\centerline{ \psfig{figure=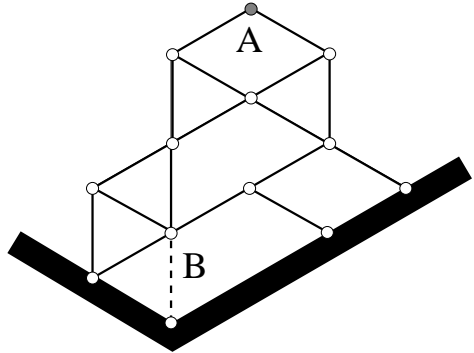,width=6cm,angle=270}}
\centerline{}
\caption{ The observed exponential growth of induced displacements with distance to
the perturbation is due to the existence of ``pantographs'' as the one shown
in this figure. Upon stretching bond B by a small amount $\delta$, site A moves
vertically by an amount $2\delta$. Conversely a unitary weight at A produces a
stress of magnitude 2 on bond B. This multiplicative effect only exists on
isostatic systems, and is lost if the network is overconstrained. In the
overconstrained case, a stretching of bond B would generate internal stresses
on other bonds that oppose the deformation, and the displacement of site A
would be much smaller. 
}
\label{fig:pantograph}
\end{figure}
Surprisingly this can be explained in very simple terms. 
The appearance of exponentially large values of displacements  is due to the
existence of  ``lever configurations'' or ``pantographs'', which amplify
displacements.  
 
Fig. \ref{fig:pantograph} shows an example of a pantograph with amplification
factor 2. When the dashed bond is stretched by $\epsilon$, site A is vertically 
shifted by $2\epsilon$. Given that there is a finite density of similar
pantographs on the system, it is clear that displacements will grow
exponentially with system height.

This amplification effect only exists in the \emph{isostatic limit}:
Pantographs as the one in Fig. \ref{fig:pantograph} are no longer effective if
blocked by overconstraints. For example, an additional (redundant) bond
between site A and the site below it would ``block'' the amplification effect
of the pantograph.  Then, a unitary stretching of bond B would induce 
stresses in the whole pantograph, but only a small displacement of site A.  \\
\begin{figure}[htb]
\vskip -0.3cm
\hbox{ \hsize=0.5\hsize
\centerline{ \psfig{figure=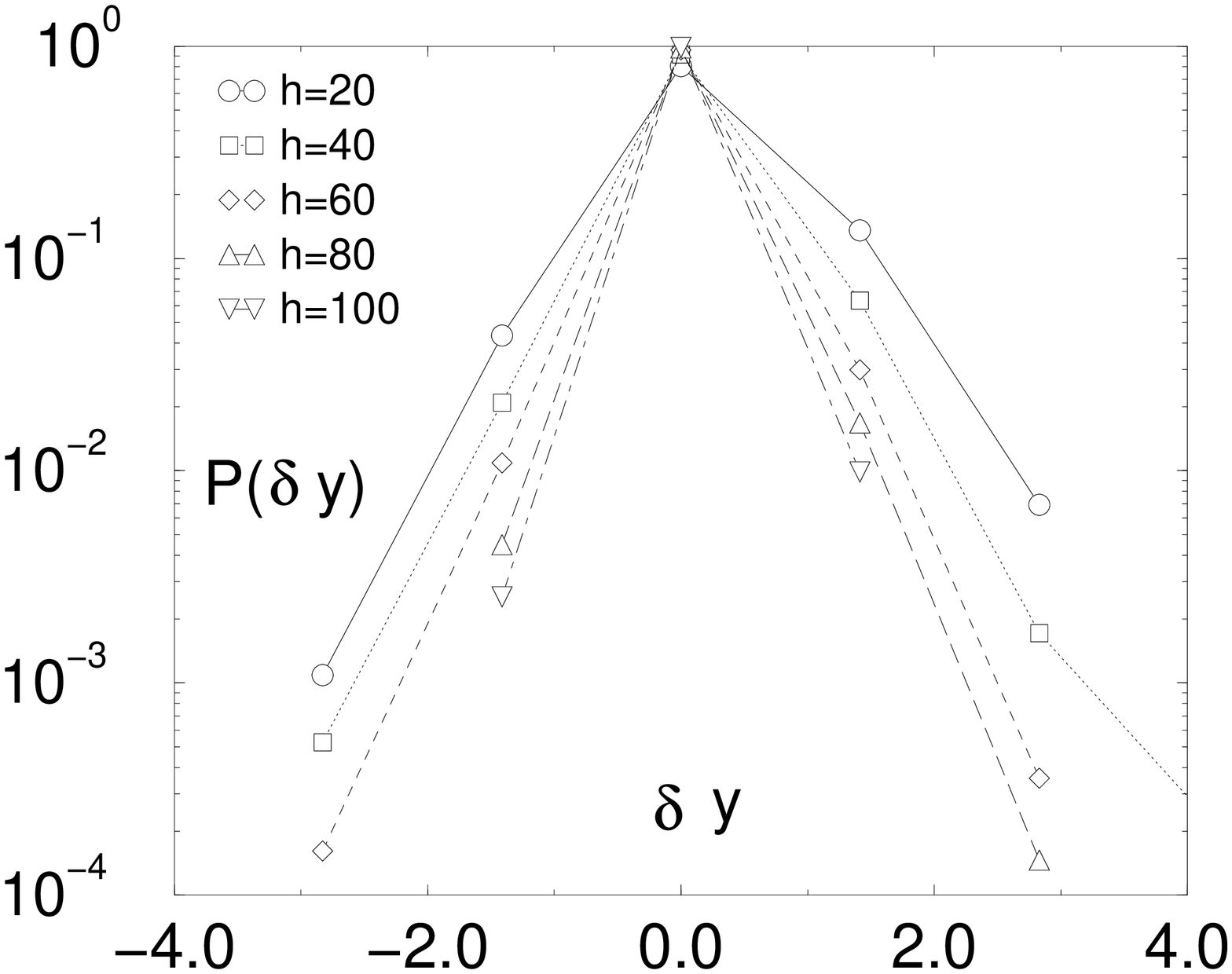,width=6cm,angle=270}}
\centerline{ \psfig{figure=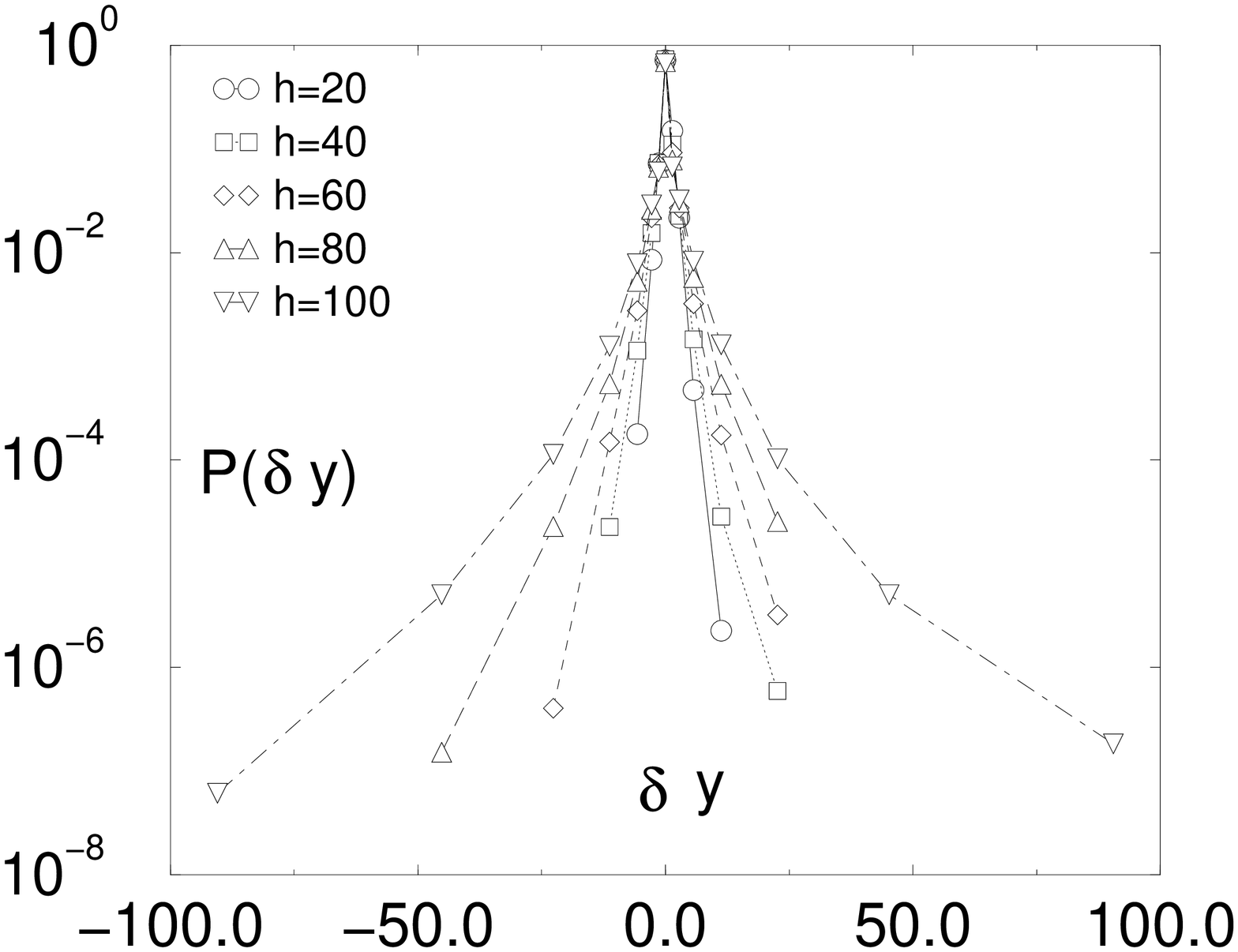,width=6cm,angle=270}}
}
\caption{ The effect of isostaticity is dramatically illustrated by a
comparison of $P_h(\delta y)$ with and without overconstraints. Here $\delta
y$ the displacement of a site, induced by a bond-length perturbation $h$
layers below it. For any nonzero  density of overconstraints (left, $O_v=0.02$
in this case), induced displacements {\it decrease} with distance. This is the
usual behavior on an elastic continuum. On the other hand if  $O_v=0$,
i.e. when the system is {\it isostatic} (right), the distribution of induced
displacements gets {\it broader} when the distance $h$ to 
the perturbation increases. This is due to the multiplicative effect of
pantographs (see Fig.~\ref{fig:pantograph}).  These results where obtained on
systems of total height 100 layers.}
\label{fig:deltaov}
\end{figure}
In order to understand why the transition occurs at zero density of
overconstraints and not at any finite density, it is extremely important to 
notice that pantographs are composed of all sites suffering displacement
when the perturbed bond is stretched. Thus a typical pantograph covers a finite
fraction of the system, and any non-zero fraction of redundant bonds
is enough to place \emph{at least} one excess bond on it, eliminating the
lever effect. This explains why anomalously large induced displacements only
exist in the isostatic limit $O_v=0$.

\section{CONCLUSIONS}
\label{sec:conclusions}

We have shown that the contact network of granular packings 
becomes \emph{exactly isostatic} in the limit of large stiffness-to-load ratio,
i.e. when the stiffness is large or the mean compressive load is small. We
have furthermore provided analytical (in 1d) and numerical (in 2d) evidence  
that isostaticity is responsible for the appearance of a large
susceptibility to perturbation, defined as the sum of the square displacements
induced by a small bond-stretching. When 
an arbitrary bond is stretched on an overconstrained system, the effect of
this perturbation is only felt locally. On the contrary, on an isostatic
system the induced displacements {\it grow} with distance. This
surprising phenomenon makes the susceptibility diverge
{\it exponentially} fast with system size, and is produced by the existence of
``pantographs'': network mechanisms that amplify displacements in the same way a
lever does.

We have also clarified the relationship between the susceptibility to
perturbation defined in this work and the experimentally observed
instability of granular networks. This was done using an equivalence between
induced displacements and the weight-stress response function. The existence
of negative values for the response function and the relation of this fact
with instability  were first discussed in the context of a phenomenological
model for stress propagation~\cite{vmodel}. In that model, the appearance of
large negative values for the response function (correctly identified as a
signature of instability by the authors) is a consequence of ad-hoc
assumptions about the way in which stresses propagate downwards. That work
though does not correctly identify the physical origin of instabilities. The
vectorial character of the transmitted quantity is {\bf not} the reason (while
it is a necessary condition), as easily illustrated by an overconstrained network.
The reason by which granular contact networks are unstable is that they are
isostatic. 

Thus stiffness produces isostaticity. Isostaticity is responsible for the
tendency to global rearrangement upon slight perturbation
of stiff granular materials. Any non-zero density of overconstraints  
is enough to destroy criticality and therefore drastically reduce
instabilities. Therefore ``soft'' granular packings are stable.  
%
%
\section*{Acknowledgments}
The author is supported by a PVE fellowship granted by CNPq, Brazil.
%
%
%
%


\begin{thebibliography}{99}
\bibitem{symmetric} 

	This value can be justified by simple geometrical
	reasoning, assuming that the radii are random variables with a small
	dispersion, and that the particles are centered on the sites of a
	regular triangular lattice.
\bibitem{HLRZ}

	I thank HLRZ J\"uelich for providing access to its supercomputing
	facilities.
\end{thebibliography}
\end{document}